\documentclass{PoS}
\usepackage{graphicx}
\usepackage[all]{xy}
\usepackage{feynmp}
\usepackage{amsmath,amssymb,bm}

\title{Pion condensation phase at finite isospin chemical potential in a Holographic QCD model}

\ShortTitle{Pion condensation phase at finite isospin chemical potential in a Holographic QCD model}

\author{\speaker{Hiroki Nishihara}$^{a}$ and Masayasu Harada,$^{a}$
\\
        \llap{$^a$}Department of Physics, Nagoya University, Nagoya 464-8602, Japan\\
        E-mail: \email{h248ra@hken.phys.nagoya-u.ac.jp},
					\email{harada@hken.phys.nagoya-u.ac.jp}}

\abstract{
We summarize main points of our recent work where we studied the pion condensation for non-zero isospin chemical potential within a holographic QCD model.
We confirmed that the second order phase transition with the mean field exponent occurs when the isospin chemical potential $\mu_I$ exceed the pion mass, which is consistent 
with the result 
obtained by the chiral effective Lagrangian at $O(p^2)$.
We find that a deviation for large $\mu_I$ region can be understood as a higher order effects in the chiral effective Lagrangian.
Our result shows that the chiral condensate defined by
$\tilde{\sigma} \equiv \sqrt{ \langle \sigma \rangle^2 + \langle \pi^a \rangle^2 }$
is almost constant in the small $\mu_I$ region, while it grows
with $\mu_I$ in the large $\mu_I$ region showing an enhancement of the chiral symmetry breaking.
}

\FullConference{XV International Conference on Hadron Spectroscopy \\
                 4-8/11/2013\\
                 Nara, Japan}

\begin{document}

\section{Introduction}
\label{sec:Introduction}

In this write-up, we summarize main points of the analysis done in Ref.~\cite{Nishihara:2014nva}, where
we studied the pion condensation phase at finite chemical potential using 
a holographic QCD model~\cite{EKSS}.

Studying QCD
at finite isospin chemical potential 
will give a clue to understand the symmetry energy which is important to describe the equation of state inside neutron 
stars.
In addition, it may give some informations on the chiral symmetry structure
of QCD.
When we turn on the 
isospin chemical potential $\mu_I$ at zero baryon number density,
the pion condensation is expected to occur at a critical point.
It was shown \cite{SS} showed that,
by using the chiral Lagrangian at $O(p^2)$,
the second order 
phase transition to the pion condensation phase 
occurs and the critical value of $\mu_I$ is equal to the pion mass.
There are so many works on the pion condensation at finite isospin chemical potential,
there are 
not many works for studying the strength of the chiral symmetry breaking.
Namely, it is interesting to ask whether or not the chiral symmetry is partially restored in the isospin matter.

In Ref.~\cite{Nishihara:2014nva}, we 
study
the pion condensation phase in a holographic QCD model~\cite{EKSS}
by introducing the mean fields for
 $\pi$, $\sigma$ and the time component of $\rho$ meson.
Our results show that
the phase transition is of the second order consistently with the one obtained in the $O(p^2)$ chiral 
Lagrangian~\cite{SS}.
It is remarkable that 
the chiral condensate defined by
$\tilde{\sigma} \equiv \sqrt{ \langle \sigma \rangle^2 + \langle \pi^a \rangle^2 }$
is almost constant in the small $\mu_I$ region, while it grows
with $\mu_I$ in the large $\mu_I$ region.
This implies that 
the chiral symmetry breaking is enhanced
by the existence of the isospin chemical potential.

\section{Model}
\label{sec:model}

In the present analysis we use 
the hard-wall holographic QCD model given in Ref.~\cite{EKSS}.
The action 
is given by
\begin{equation}
S_5=\int d^4x \int_\epsilon^{z_m} dz\sqrt{g} \mathrm{Tr} \left[ |DX|^2 +3|X|^2 -\frac{1}{4g_5^2}\left(F^2_L+F^2_R\right)\right]
+\mathcal{L}^{BD}_5
\ ,
\label{action}
\end{equation}
where the metric is $ds^2=a^2(z) \left(\eta_{\mu\nu}dx^\mu dx^\nu -dz^2\right)$ with $a(z)=1/z$ and the fifth direction $z$ has the UV and the IR cutoffs, $\epsilon$ and $z_m$.
This theory has the chiral symmetry U(2$)_{\rm L}\times$U(2)$_{\rm R}$, and that is consisted by a scalar field $X$ and the
gauge fields corresponding to the chiral symmetry U(2$)_{\rm L}\times$U(2)$_{\rm R}$
The action contains the IR boundary term $\mathcal{L}^{BD}_5$ and 
the profile of $\mathcal{L}^{BD}_5$ is the $\phi$ fourth type of the $X$ field which has two parameters $\lambda$ and $m^2$.
In the following analysis we adopt the $L_5=R_5=0$ gauge, and use the IR-boundary 
condition $F^L_{5\mu}|_{z_m}=F^R_{5\mu}|_{z_m}=0$.

In the vacuum 
the
chiral symmetry is spontaneously broken down to U$(2)_V$ by the vacuum expectation value of $X$. 
This is given by solving equation of motion and the solution of the $X$ has two parameters $m_q$ and $\sigma$,  
where $m_q$ corresponds to the current quark mass and $\sigma$ to 
the quark condensate~\cite{EKSS}.
They are related with $\lambda$ and $m^2$ 
by the IR-boundary condition of the $X$. 
A parameter $g_5^2$ is determined by matching with QCD as
$
g_5^2=\frac{12\pi^2}{N_c}. 
$
The pion is described as a linear combination of the lowest eigenstate of 
$\pi^a$ and the longitudinal mode of $A^a_\mu$,  and the $\rho$ meson is the lowest eigenstate of $V_\mu$. 
The values of the $m_q$ and $z_m$ 
together with that of 
the relation between these parameters $\lambda$ and $m^2$ 
are fixed by fitting them to the pion mass  $m_\pi=139.6$ MeV, the $\rho$ meson mass $m_\rho=775.8$ MeV and the pion decay constant $f_\pi=92.4$ MeV: 
$
m_q = 2.29\,\mbox{MeV}\ , 
z_m = 1/(323\,\mbox{MeV}) 
\ .
\label{parameter1}
$
In the present analysis, we use the $a_0$ meson mass $m_{a_0}=980$ MeV
as a reference value, which fixes $m^2 = 5.39$ 
and $\lambda=4.4$, and see the dependence of our results on the scalar
meson mass.

\section{Pion condensation phase}
\label{sec:PCP}

In this section we study the pion condensation for finite isospin chemical potential $\mu_I$
in the holographic QCD model introduced in 
section~\ref{sec:model}.

The isospin chemical potential $\mu_I$ is introduced
as a UV-boundary value of the time component of 
the gauge field of SU(2)$_{\rm V}$ symmetry as
\begin{eqnarray}
V^3_0(z)|_\epsilon=\mu_I \ ,
\label{mu}
\end{eqnarray}
where the superscript $3$ indicates the third component of the isospin corresponding to the neutral $\rho$ meson.
Here we study 
the pion condensation phase 
for small $\mu_I$,
then we assume that the rotational 
symmetry O(3) is not broken by e.g. the $\rho$ meson condensation:$L_i=R_i=0$.
 We also assume the time-independent condensate, 
then the vacuum structure is determined by studying the mean fields
of five-dimensional fields which do not depend on the four-dimensional coordinate.
Furtheremore,
we also take the mean fields for
the neutral pion, the iso-triplet scalar meson ($a_0$ meson) and the iso-singlet pseudoscalar meson ($\eta$ meson) 
to be zero
 in the pion condensation phase .

\begin{figure}[ht]
 \begin{center}
  \includegraphics[width=55mm]{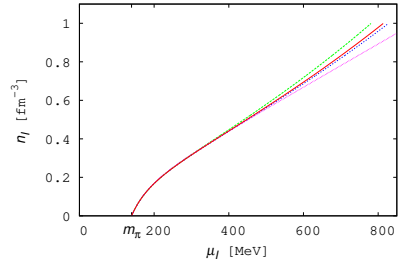}
 \end{center}
 \caption[]{
Relation between the
isospin number density $n_I$ and the isospin number chemical potential $\mu_I$.
The green, red and blue  curves show our results for $\lambda = 1$, $4.4$ and $100$,
respectively.
The pink dashed-curve shows the result given by the chiral Lagrangian in Ref.~\cite{SS}. 
Each choice of $\lambda$ corresponds to $m_{a_0}=610$MeV, $980$MeV
and $1210$MeV, respectively.
}
 \label{fig nI vs mu}
\end{figure}

We show the resultant relation between $\mu_I$
 and the 
isospin density in Fig.~\ref{fig nI vs mu}
for $\lambda = 1$, $4.4$ and $100$ corresponding to
$m_{a_0}=610$MeV, $980$MeV
and $1210$MeV.
This shows that the phase transition is of the second 
order and the critical chemical potential is predicted to be equal to 
the pion mass.
This is consistent with the result obtained by the chiral Lagrangian approach
in Ref.~\cite{SS}.
Furthermore,
our result on 
the relation between isospin number density and isospin chemical potential 
for small $\mu_I$ 
agrees with the following one obtained by O$(p^2)$ chiral Lagrangian~\cite{SS}:
\begin{eqnarray}
n_I&=&f_\pi^2 \mu_I\left(1-\frac{m_\pi^4}{\mu_I^4}\right).
\label{iso-density SS}
\end{eqnarray}
For $\mu_I > 500$\,MeV, there is a difference between our predictions
and the one from O$(p^2)$ chiral Lagrangian, which can be understood as
the higher order contribution as we will show in the next section.

\begin{figure}[ht]
 \begin{minipage}{0.47\hsize}
  \begin{center}
 	 \includegraphics[scale=0.45]{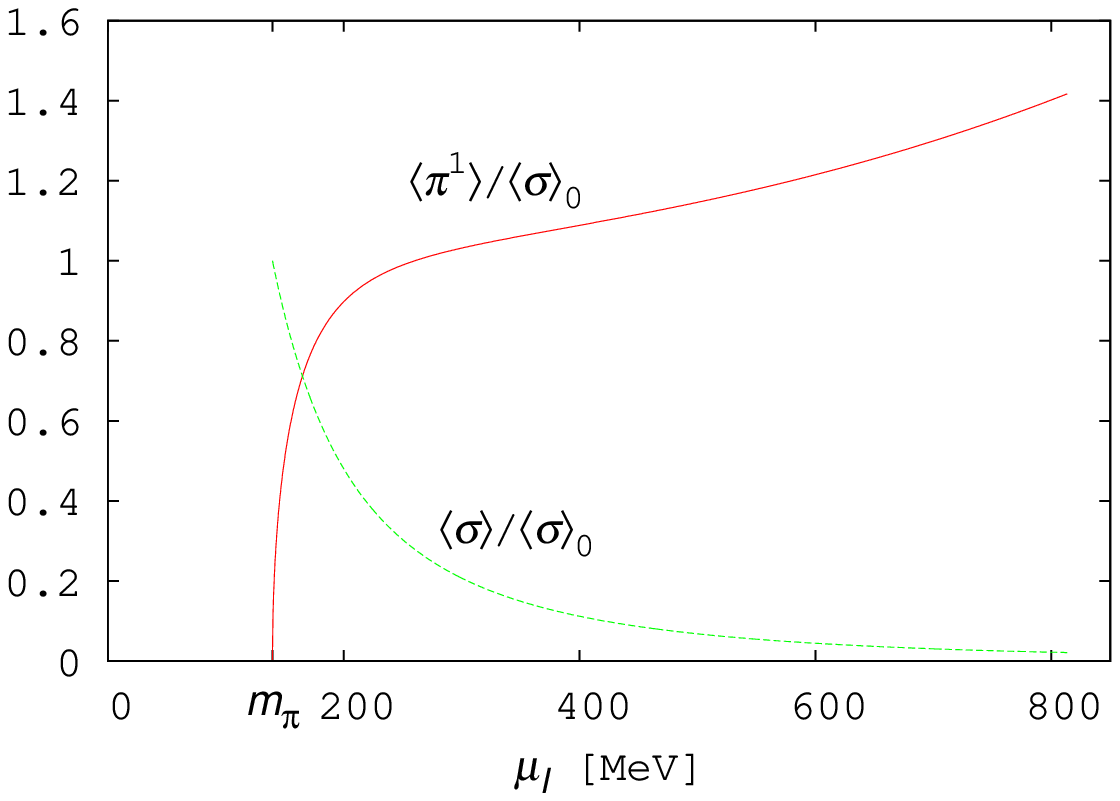}
 \caption{
$\mu_I$ dependences of the $\pi$-condensate 
(red curve) and
the ``$\sigma$"-condensate (green curve). 
}
 \label{fig v vs mu}
  \end{center}
 \end{minipage}
\hspace{0.2cm}
\begin{minipage}{0.5\hsize}
\end{minipage}
\begin{minipage}{0.47\hsize}
  \begin{center}
   \includegraphics[scale=0.45]{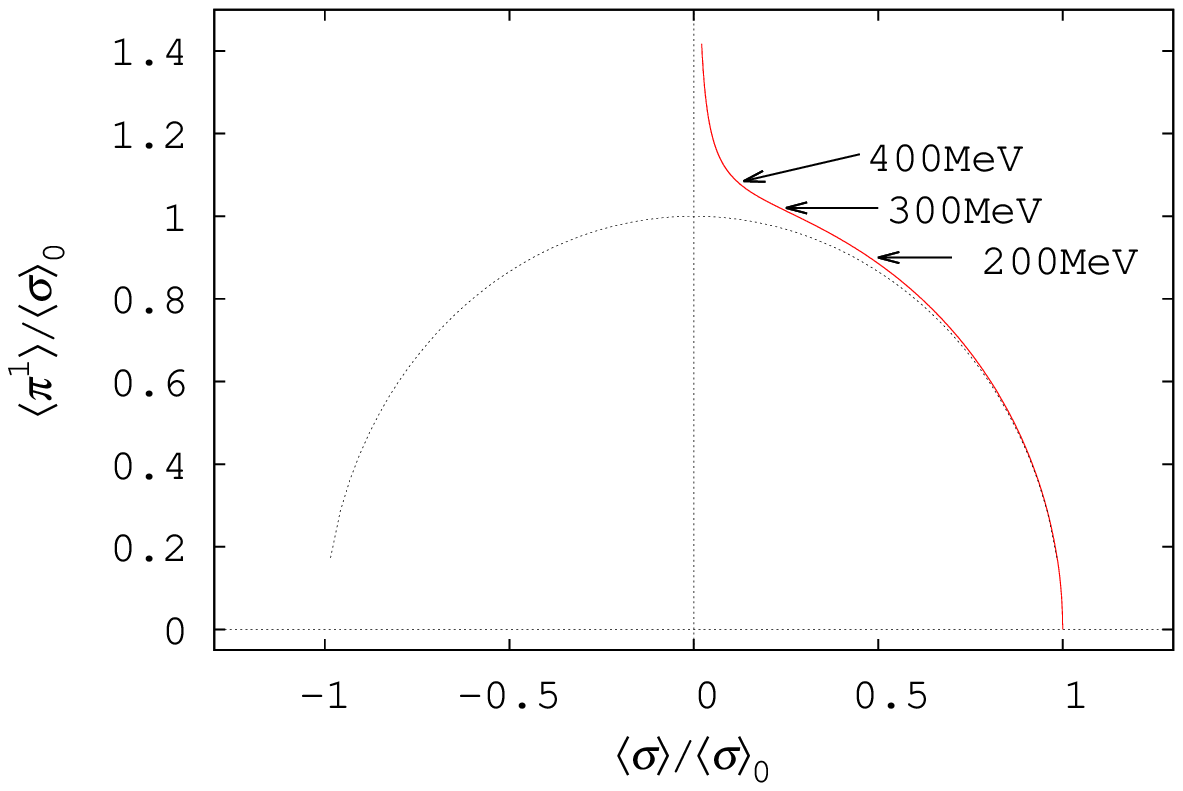}
 \caption{The chiral circle is showed as the red curve.  The black curve is an unit circle.}
 \label{fig chiral circle}
  \end{center}
 \end{minipage}
\end{figure}

We show the $\mu_I$ dependences of these condensate in Fig.~\ref{fig v vs mu} ,
where 
$\langle\sigma\rangle_0$ is the ``$\sigma$"-condensate at $\mu_I=0$.
This shows that 
the ``$\sigma$''-condensate decreases rapidly after the phase transition where the $\pi$-condensate grows rapidly.
The ``$\sigma$"-condensate becomes very small for $\mu_I \gtrsim 400$\,MeV, 
while the 
$\pi$-condensate keeps increasing.
Using the form $\langle\pi^a\rangle \propto \left(\mu_I-\mu_I^c\right)^\nu$ near the
phase transition point, we fit 
the critical exponent $\nu$ to obtain 
$\nu=\frac{1}{2}$.
This implies that 
the phase transition here is the mean field type.

We also show
the ``chiral circle"  in 
Fig.~\ref{fig chiral circle}. 
It is remarkable that
the value of the ``chiral condensate" defined by
\begin{equation}
\tilde{\langle \sigma\rangle} = \sqrt{ \langle \sigma \rangle^2 + \langle \pi^a \rangle^2 }
\end{equation}
is constant for increasing isospin chemical potential $\mu_I$ for $\mu_I \lesssim 300$\,MeV,
and that it grows rapidly in the large $\mu_I$ region.

\section{Comparison with the chiral Lagrangian}
\label{sec:chiral}

In this section, we compare our result for the relation between the
isospin number density $n_I$ and  $\mu_I$ shown in 
Fig.~\ref{fig nI vs mu} as well as the $\mu_I$-dependences of the $\pi$-condensate and the ``$\sigma$''-condensate in Fig.~\ref{fig v vs mu}, 
with the ones from the chiral Lagrangian including the O($p^4$) terms.
Here we use the
chiral Lagrangian for two flavor case~\cite{GL}.
One can simply introduce
the isospin chemical potential $\mu_I$ as vacuum expectation values of 
these external gauge fields as
$\langle {\mathcal L}^3_\mu \rangle = \langle {\mathcal R}^3_\mu \rangle 
= \frac{\mu_I}{2} \delta_{0\mu}$~\cite{SS}.
By minimizing the effective potential, we can obtain the relation between $n_I$ and $\mu_I$,
 $\frac{\langle \sigma \rangle}{\langle \sigma \rangle_0}$ and 
$\frac{\langle \pi^1 \rangle}{\langle \sigma \rangle_0}$.

In Ref.\cite{Nishihara:2014nva},
we fitted the values of the relevant low-energy constants in the chiral Lagrangian 
to our result on the $\mu_I$ dependence of $n_I$
, $\frac{\langle \sigma \rangle}{\langle \sigma \rangle_0}$ and 
$\frac{\langle \pi^1 \rangle}{\langle \sigma \rangle_0}$.
We find that the deviation of our result from the one 
obtained from the $O(p^2)$ chiral Lagrangian is actually explained by 
including effects of $O(p^4)$ terms.

\section{A summary and discussions}
\label{sec:summary}

We studied the phase transition to the pion condensation phase for 
finite isospin chemical potential
using the holographic QCD model given in Refs.~\cite{EKSS},
by introducding
the isospin chemical potential $\mu_I$ 
as a UV-boundary value of the time component of 
the gauge field of SU(2)$_{\rm V}$ symmetry as
$V^3_0(z)|_\epsilon=\mu_I $.
We assumed non-existence of vector meson condensates since we are interested
in studying the small $\mu_I$ region.
Furthermore, we assumed that the neutral pion does not condense.
We solved the coupled equations of motion for the $\pi$-condensate and
``$\sigma$''-condensate together $V^3_0$ to determine $\mu_I$ as an eigenvalue.

Our result shows that the phase transition is of the second 
order and the critical chemical potential is predicted to be equal to 
the pion mass.
This is consistent with the result obtained by the chiral Lagrangian approach
in Ref.~\cite{SS}, but contrary to the result in Ref.~\cite{Albrecht:2010eg}. 
Furthermore,
our result on 
the relation between isospin number density and isospin chemical potential 
for small $\mu_I$ 
agrees with the following one obtained by O$(p^2)$ chiral Lagrangian~\cite{SS}.
For large $\mu_I$ ($> 500$\,MeV), 
there is a difference between our predictions
and the one from O$(p^2)$ chiral Lagrangian, which is shown to be  understood as
the O$(p^4)$  contributions.

We also studied the $\mu_I$ dependence of the $\pi$-condensate and ``$\sigma$''-condensate.
Our result shows that, at the phase transition point, 
the $\pi$-condensate increases from zero with the mean field exponent 
Furthermore, we find that
the ``$\sigma$''-condensate decreases rapidly after the phase transition where the $\pi$-condensate grows rapidly, while 
the value of the "chiral condensate" defined by
$
\tilde{\langle \sigma\rangle} = \sqrt{ \langle \sigma \rangle^2 + \langle \pi^a \rangle^2 }
$
is constant for $\mu_I \lesssim 300$\,MeV,
and that it grows rapidly in the large $\mu_I$ region.
This indicates that the chiral symmetry restoration at finite baryon density and/or finite temperature will be delayed when non-zero isospin chemical potential is turned on.

\section*{Acknowledgements}

We would like to thank Shin Nakamura for useful discussions and comments.
This work 
was supported in part by Grant-in-Aid for Scientific Research
on Innovative Areas (No. 2104) ``Quest on New
Hadrons with Variety of Flavors'' from MEXT,
and by the JSPS Grant-in-Aid for Scientific Research
(S) No. 22224003, (c) No. 24540266.


\begin{thebibliography}{100}

\bibitem{Nishihara:2014nva} 
  H.~Nishihara and M.~Harada,
  arXiv:1401.2928 [hep-ph].

\bibitem{EKSS}
	J. Erlich et al,
	Phys.\ Rev.\ Lett. {\bf 95} (2005) 261602,~
	L. D. Rold and A. Pomarol, 
	Nuclear Physics B {\bf 721} (2005) 79-97,~
	L. D. Rold and A. Pomarol, 
	JHEP {\bf 0601} (2006) 157 

\bibitem{SS}
	D. T. Son and M. A. Stephanov, 
	Phys.\ Rev.\ Lett. {\bf 86} (2001) 592-595 

\bibitem{GL}
	J. Gasser and H. Leutwyler,
	Nuclear Physics B {\bf 250} (1985) 465-516,~
  M.~Harada and K.~Yamawaki,
  Phys.\ Rept.\  {\bf 381}, 1 (2003).

\end{thebibliography}
\end{document}